\documentclass[journal=jpclcd,manuscript=article]{achemso}

\usepackage[version=3]{mhchem} 
\usepackage[T1]{fontenc}       
\usepackage{graphicx}
\usepackage{amsfonts}
\usepackage{color}
\usepackage{amsmath}
\usepackage{amssymb}

\def\<{\langle}
\def\>{\rangle}
\def\tw{t_w}

\newcommand{\textgx}[1]{\textcolor{black}{#1}}

\author{Raffaele Pastore}
\affiliation{
CNR--SPIN, sezione di Napoli,
Dipartimento di Fisica, Campus universitario di Monte S. Angelo, Via Cintia, 80126 Napoli, Italy
}
\email{raffaele.pastore@spin.cnr.it}

\author{Giuseppe Pesce}
\affiliation{
Dipartimento di Fisica, Universit\'a di Napoli Federico II, Campus universitario di Monte S. Angelo, Via Cintia, 80126 Napoli, Italy
}

\author{Antonio Sasso}
\affiliation{
Dipartimento di Fisica, Universit\'a di Napoli Federico II, Campus universitario di Monte S. Angelo, Via Cintia, 80126 Napoli, Italy
}

\author{Massimo Pica Ciamarra}
\affiliation{
Division of Physics and Applied Physics, School of Physical and Mathematical Sciences, Nanyang Technological University, Singapore
}
\alsoaffiliation{
CNR--SPIN, sezione di Napoli,
Dipartimento di Fisica, Campus universitario di Monte S. Angelo, Via Cintia, 80126 Napoli, Italy
}

\title{Cage Size and Jump Precursors in Glass-Forming Liquids: Experiment and Simulations}

\begin{document}

\begin{tocentry}

\begin{center}
\includegraphics*[width=5.0cm]{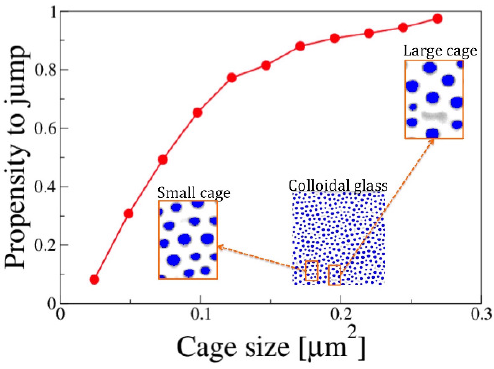}
\end{center}

\end{tocentry}

\begin{abstract}

Glassy dynamics is intermittent,
as particles  suddenly jump out of the cage formed by their neighbours, and
heterogeneous, as these jumps are not uniformly distributed across the system.
Relating these features of the dynamics to the diverse local environments
explored by the particles is essential to rationalize the relaxation process.
Here we investigate this issue characterizing the local environment of a particle with the amplitude of its short time vibrational motion,
as determined by segmenting in cages and jumps the particle trajectories. 
Both simulations of supercooled liquids and experiments on colloidal suspensions show 
that particles in large cages are likely to jump after a small time-lag, and that, on average, the cage enlarges shortly before the particle jumps.
At large time-lags, the cage has essentially a constant value, which is smaller for longer-lasting cages. 
Finally, we clarify how this coupling between cage size and duration controls the average behaviour and
opens the way to a better understanding of the relaxation process in \textgx{glass--forming liquids}. 
 \end{abstract}

\textgx{Molecular liquids on lowering the temperature and colloidal suspensions on increasing the volume fraction
exhibit a glass transition from a liquid-like to an amorphous solid-like state~\cite{Nagel, Trappe, Solomon, Debenedetti, Ngai, Xu}.
On approaching this transition,  the dynamics becomes intermittent and shows large spatio-temporal fluctuations,
also known as dynamic heterogeneities~\cite{DHbook, Paluch}.
These dynamic features are currently emerging as a common hallmark of many complex systems,
such as foams, gels~\cite{foam} and fiber networks~\cite{fiber} as well as biological materials, such as cell tissues~\cite{Fredberg, Cerbino_cells} and microswimmers~\cite{active},
in which primary particles move in a crowded environment. 
As a consequence, there is a great deal of interest in applying concepts developed by the glass community to understand the behavior of these systems.
As telling example, recent results show that the degree of dynamic heterogeneities highlight 
and allow rationalizing pathological conditions in epithelial cell tissues.~\cite{Fredberg, Cerbino_cells}
}
 
In glass-forming liquids, the dynamics is spatio-temporal heterogeneous since the probability that a particle rearranges
in a given time interval is not spatially uniform,
as long as the considered time interval is smaller than the relaxation time. 
Since the local environment of a particle affects its short time motion,
dynamic heterogeneities indicates the presence of structural heterogeneities.
This observation triggered a resurgence of interest~\cite{royall2015,tanaka2014} in the search
of connections between structure and dynamics
in supercooled liquids, a notorious difficult task.
This problem can be somehow simplified assuming the structure to influence the short time dynamics,
and the short time dynamics to influence the relaxation process. Thus, instead of looking for
connections between structure and long time dynamics, one looks for connections between short time
dynamics and relaxation. Research in this direction clarified that particles highly mobile
on a short time scale are also those that most probably 
will undergo a significant displacement on the structural relaxation timescale. 
Operatively, particles highly mobile on a short time scale can be identified,
somehow equivalently, as those located where soft vibrational modes are localized~\cite{Widmer-Cooper, Manning, Delgado}, 
as those in regions with small local elastic constants~\cite{Barrat}, as those with a large free volume\cite{Starr}, 
and  as those having a large vibrational motion~\cite{Candelier, Samwer}.
While the existence of an interplay between short time dynamics and structural relaxation is clear~\cite{Leporini},
quantitative relations between these two features, at the single particle level, are still lacking. 

Here we tackle this issue through a combined experimental and numerical study of two popular fragile glass-forming models,
we have investigated in previous works~\cite{SM14, SM15, SM15_corr, JSTAT16}. 
Briefly, we perform {\it i)} experiments on a nearly two-dimensional suspension of hard-sphere colloids,
whose dynamics slows down on increasing the volume fraction, and, 
{\it ii)} molecular dynamic simulations of a two dimensional system of soft disks,
whose dynamics slows down on lowering the temperature (see Methods for details on the investigated systems). 
By taking advantage of the intermittent cage-jump motion  characterizing the single-particle dynamics
in supercooled liquids and glasses~\cite{Weeks, Chauduri, SM_review, Voll, Baschnagel, Wales, Bernini, Schweizer}, we segment the particle trajectories in cage and jumps~\cite{SM14} 
and use the cage size as a proxy of the short time motion of a particle; similarly, 
we use jumps as proxies of the local relaxation.

As a nearly instantaneous measure of the cage size at time $t$, we use the fluctuations $S^2_C$ of the
particle position in a short time interval $[t-\delta t/2, t+\delta t/2]$, with $\delta t$ of the order of the
Debye-Waller factor characteristic time~\cite{Leporini}.
Further details on the cage--jump detection algorithm and on the cage size estimate are discussed in Methods.
For  a particle in a cage of size $S^2_C$ at time $t$, we consider the conditional probability distributions,
$P_{CC}(S^2_C,\Delta t)$, that the particle stays permanently caged up to $t+\Delta t$, 
and, $P_{CJ}(S^2,\Delta t)$, that the particle starts jumping at time $t+\Delta t$.  
These conditional probabilities do not depend on $t$ as the systems we consider are in thermal equilibrium,
and, therefore, invariant under time translations. For every $\Delta t$, the conditional probabilities are normalized,
$\int P_{CC}(S^2,\Delta t) dS^2 = \int P_{CJ}(S^2,\Delta t) dS^2 = 1$.

The influence of the cage size at time $t$ on the probability that a particle will start jumping at later time $t+\Delta t$
can be quantified introducing a jump propensity, $\Pi_J(S^2_C,\Delta t)$, 
\begin{equation}
\label{eq:propensity2} 
\Pi_J(S^2_C,\Delta t) = \frac{P_{CJ}(S^2_C,\Delta t)}{P_{CC}(S^2_C,\Delta t)+P_{CJ}(S^2_C,\Delta t)}.
\end{equation}
When the probability that a particle jumps is not correlated to the size of its cage, 
$\Pi_J(S^2_C,\Delta t)) = \Pi_{\rm dec} = 1/2$. Conversely, $\Pi_J(S^2_C,\Delta t)) \simeq 0$
and $\Pi_J(S^2_C,\Delta t)) \simeq 1$ indicate that particles with cages of size $S^2_C$ are very
unlikely and very likely to jump after a time $\Delta t$, respectively.

{\it Jump propensity at short times.}
To investigate how the cage size correlates with the probability that a particle is about to jump,
we start by considering 
a time-lag, $\Delta t = \delta t$,
that is the smallest time to probe $S^2_C(t)$ and $S^2_C(t+\Delta t)$ over non-overlapping time windows,
according to our cage-jump algorithm.
\begin{figure}[t!!]
\begin{center}
\includegraphics*[scale=0.33]{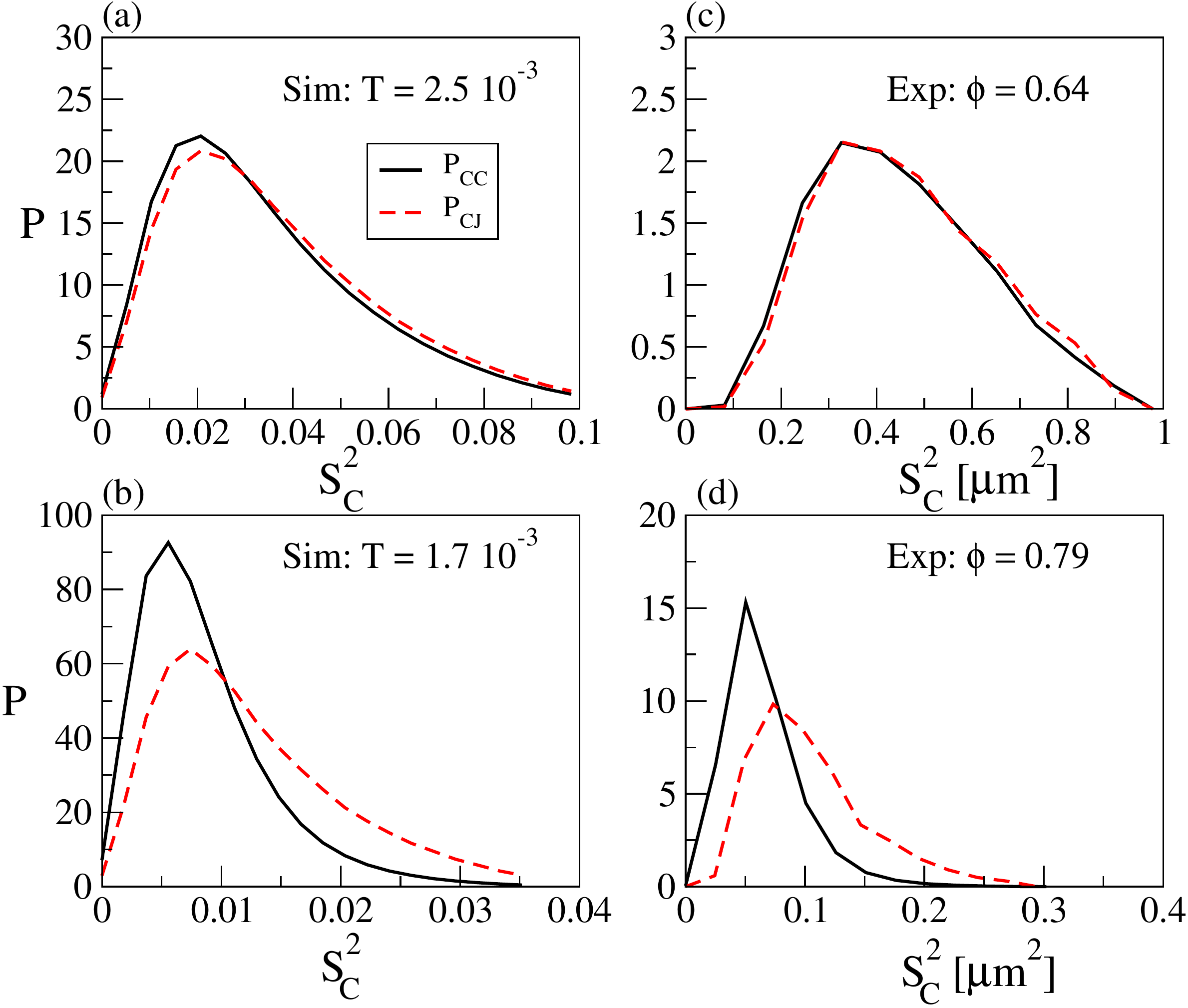}
\end{center}
\caption{Conditional probability distributions, $P_{CC}(S^2_C,\Delta t)$ and  $P_{CJ}(S^2_{CJ},\Delta t)$,
as a function of the cage size $S^2_C$ and at a short time-lag, $\Delta t=\delta t$.
Panel a and b: simulations at the lowest and the highest investigated temperature, respectively.
Panel c and d: experiments at the largest and the smallest investigated volume fraction, respectively.
\label{fig:c_j}}
\end{figure}
Physically, we are investigating correlations separated by a timescale fixed by the Debye-Waller factor time.
Fig.~\ref{fig:c_j} illustrates our numerical and experimental results for the conditional probability distributions $P_{CC}$ and $P_{CJ}$.
In the numerical simulations, we observe the two distributions to be 
almost indistinguishable at high temperature (panel a).
Conversely, clear differences emerge at low temperature (panel b), 
$P_{CJ}$ having a much fatter tail than $P_{CC}$.
Analogous results are observed in the experiment, when comparing the distributions
measured at low and at high volume fractions, as in panels c and d.

\begin{figure}[t!!]
\begin{center}
\includegraphics*[scale=1.0]{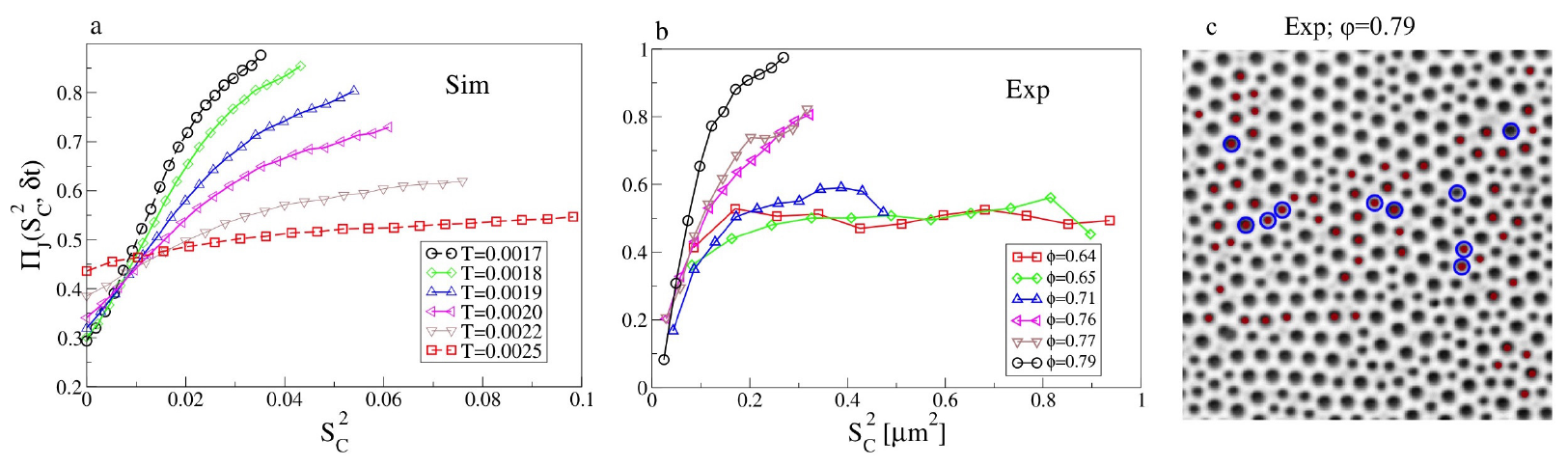}

\end{center}
\caption{Dependence of the short-time jump propensity, $\Pi_J(S^2_C,\delta t)$, on the cage size, in the liquid and in the supercooled regime,
in numerical simulations (a), and in the experiment (b). 
c) Snapshot of the investigated colloidal suspension at volume fraction $\phi=0.79$ , taken at a time $t=0 s$.
Full red circles indicate caged particles with a large vibrational motion, ($S^2_C(t)>0.1 \mu m^2)$, 
while empty circles surround the caged particles that will jump in the time interval $[30 s,150 s]$.
\label{fig:propensity}}
\end{figure}

This effect can be further quantified by the jump propensity, $\Pi_J(S^2_C,\delta t)$, of Eq.~\ref{eq:propensity2}.
The dependence of this propensity on the cage size is illustrated in Fig.~\ref{fig:propensity}. Panel a
shows results for the numerical system at different temperatures, while panel b shows experimental results
at different volume fractions. At high temperatures or low volume fraction, i.e. in the conventional liquid phase, no correlations are expected between the 
cage amplitude and the jumping ability of a particle, and indeed we do observe $\Pi_J(S^2_C,\delta t) \simeq \Pi_{\rm dec} = 1/2$.
Conversely, in the supercooled regime $\Pi_J(S^2_C,\delta t)$ is a growing function of $S^2_C$, indicating that
the larger the cage of a particle, the more likely the particle will jump after a short delay.
It is worth noticing that, even for the most supercooled systems, the propensity is close to unity only for the largest cage size detected,
whereas it seems to saturates at a progressively lower values as supercooling becomes more moderate.

From these figures we learn that, in the supercooled regime, particles with a large cage are more likely to
start jumping after a short time-lag, although jumps originating from small cages are still possible.
Figure~\ref{fig:propensity}c provides a direct visualization of this effect for the colloidal systems we have investigated. 
The figure is a snapshot of the system in a deeply supercooled state (the highest investigated volume fraction) and
highlights that a large fraction of the jumps, occurring shortly after the considered frame, do originate from large cages. 
However, there also a few jumps originating from small cages, as well as, many large cages not giving rise to a jump on the considered timescale.

\begin{figure}[t!!]
\begin{center}
\includegraphics*[scale=1.0]{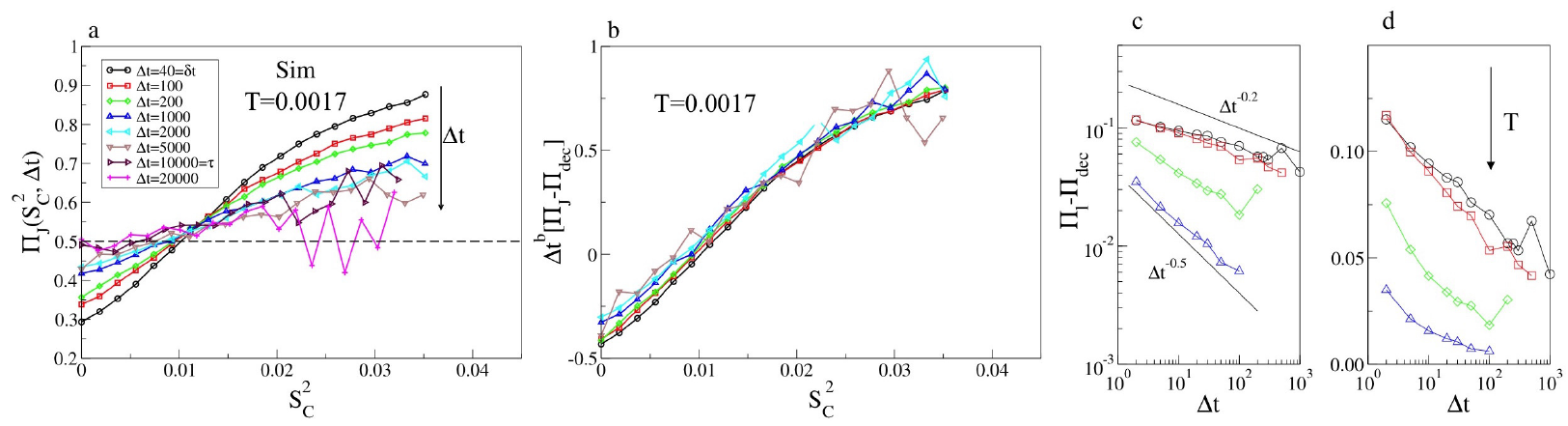}
\end{center}
\caption{
Panel a illustrates the dependence of the jump propensity on the cage size, at different time-lag $\Delta t$, as indicated.
Data corresponding to different $\Delta t$ can be collapsed on the same master functional form (see Eq.~\ref{eq:scaling}),
 as in panel b, suggesting a slow power-law decay of the temporal correlations. Data at the longest time-lags are too noise and
 are not reported in this scaling. 
 Panel c illustrates the excess jump propensity averaged over the 30\% largest cages, $\Pi_l-\Pi_{dec}$, as a function of the time-lag, $\Delta t$, for simulations at different temperature.
From top to bottom: $T=1.7~10^{-3}, 1.8~10^{-3}, 1.9~10^{-3}, 2.0~10^{-3}$. Solid line indicates that the decay is compatible with a power law, the exponent decreasing on cooling, as indicated.
Panel d reports the same data in a lin-log plot, clarifying that, at low temperature, the decay is also consistent with a logarithmic law.  
\label{fig:propensitydt}}
\end{figure}

{\it Jump propensity relaxation.}
In the limit in which $\Delta t$ is much larger than the relaxation time, the probability that a particle
jumps at time $t +\Delta t$ should not depend on the size of its cage at time $t$. Thus,
for larger $\Delta t$ one expects $\Pi(S^2_c,\Delta t) = \Pi_{\rm dec} = 1/2$.
Here we consider the relaxation dynamics of the propensity, $\Pi_J(S^2_c,\Delta t)$, by focusing on its $\Delta t$ dependence.
Fig.~\ref{fig:propensitydt}a illustrates numerical results obtained at low temperature, 
the experimental ones at high volume fraction being \textgx{similar}.
As $\Delta t$ increases, the propensity $\Pi(S^2_C,\Delta t)$ evolves and approaches $\Pi_{\rm dec} = 1/2$, as expected.
Fig.~\ref{fig:propensitydt}b supports the following scaling relation form for
the $\Delta t$ dependence of the propensity,
\begin{equation}
\label{eq:scaling}
\Pi(S^2_C,\Delta t) = \Pi_{\rm dec}+\Delta t^{-b} f(S^2_C)
\end{equation}
with $f(x)$ a universal scaling function, and $b \simeq 0.2$. 
Thus, the decay of the jump propensity 
occurs through a slow power-law process. 

The same decay process is seen to
occur at other temperatures, with the exponent $b$ increasing with $T$.
This is shown in Fig.~\ref{fig:propensitydt}c, that reports the $\Delta t$
dependence of $\Pi_l(\Delta t)-\Pi_{\rm dec}$, $\Pi_l$ being the propensity averaged over the largest cages (30\%).
In panel d, we illustrate the same data on a lin--log scale, to clarify that,
at low temperature, a logarithmic behaviour also describes this decay.

{\it Cage dynamics.}
Since the jump propensity depends on the cage size, 
the jump of a particle might be preceded by an enlarging of the cage,
possibly driven by changes in the local structure.
Here we consider this issue investigating the time evolution of the cage size.

For each cage of duration $t_{w,i}$ and end time $t_{J,i}$ (where a jump starts),
we monitor the cage size $S^2_{C,i}(\Delta t)$ as a function of the time left before the jump, 
$\Delta t = t_{J,i}-t$, with $t\in[t_{J,i}-t_{w,i},~ t_{J,i}]$.
We first consider the average cage size of all particles that start a jump after a same time-lag $\Delta t$:
\begin{equation}
\label{eq:S2_total}
\<S^2_C(\Delta t)\>= \frac{\sum_i S^2_{C,i}(\Delta t) \Theta(\Delta t-t_{w,i})}{\sum_i\Theta(\Delta t-t_{w,i})}
\end{equation}
where the sums runs over all the detected cages and the Heaviside function, $\Theta$, accounts for the fact
that, at a given $\Delta t$, only cages with $t_{w,i}>\Delta t$ do contribute to the average.

\begin{figure}[!!t]
\begin{center}
\includegraphics*[scale=1.0]{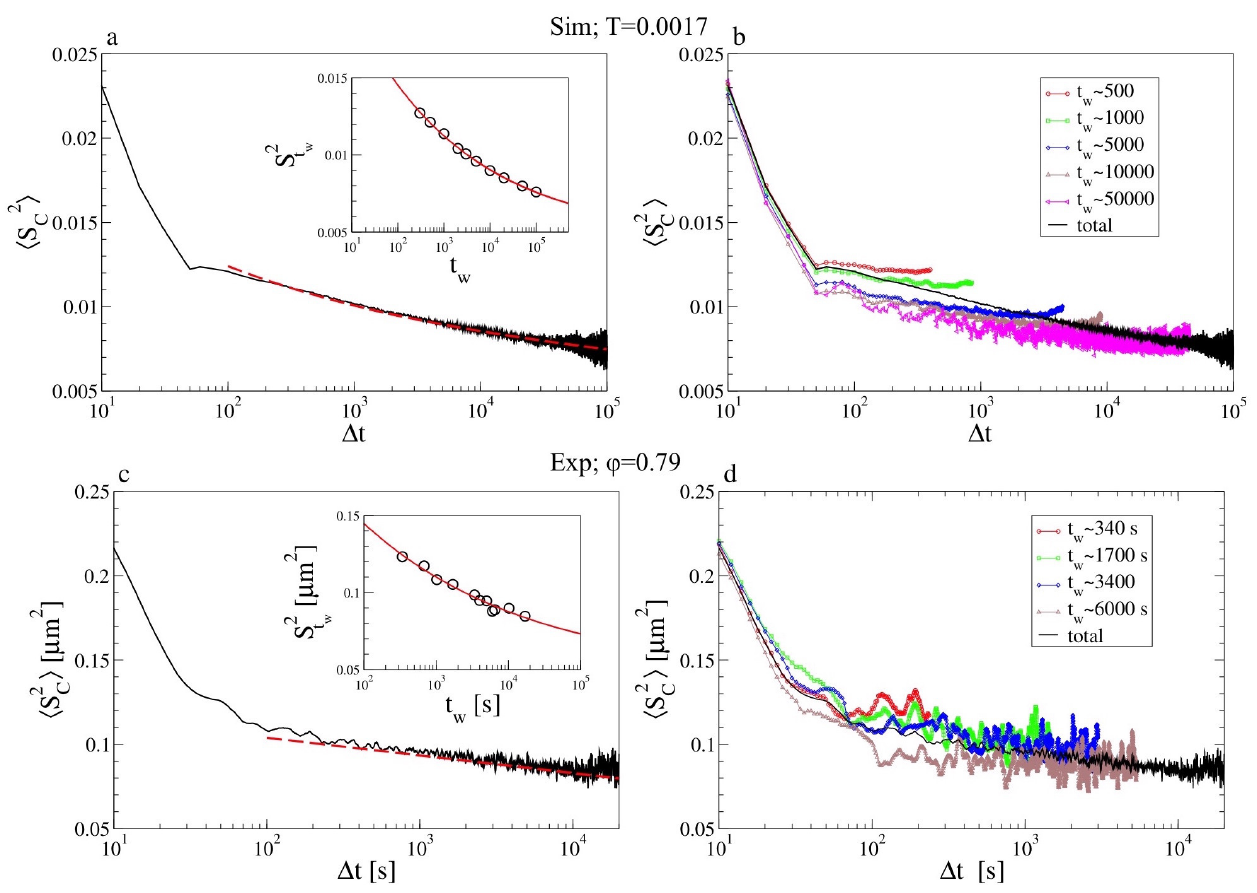}
\end{center}
\caption{From simulations at the lowest investigated temperature a)  cage size averaged over the total ensemble of cages,$\<S^2_C(\Delta t)\>$ (solid line). 
At long time,  data are correctly approximated by numerically computing the model of Eq.~\ref{eq:final} (dashed line),
using a stretched exponential fit for $P(t_w)$~\cite{JSTAT16} and the fit for $S^2_{t_w}$ as reported in the inset. 
b) Cage size averaged over the total ensemble of cages (line) and over sub-ensembles of cages
of fixed lifetimes $t_w$ (line-points).
Upper Inset: $S^2_{t_w}$ as a function of $t_w$, as obtained by a constant fit to the long-time behaviour of  $\<S^2_C(\Delta t)\>_{t_w}$. 
The solid line is a fit to the data , $S^2_{t_w} = A/t_w^\beta + S_{\infty}$, with $A=0.02\pm0.001$, $\beta=0.18\pm0.01$, $S_{\infty}=0.0047\pm0.0001$.
Panels c), d) and the lower Inset reported the same analysis for experiments at the largest investigated volume fraction.
The fitted parameters in the inset are $A=0.23\pm0.01$, $\beta=0.19\pm0.01$, $S_{\infty}=0.047\pm0.001$
\label{fig:cage_t_tw}}
\end{figure}

Figure~\ref{fig:cage_t_tw}a shows $\<S^2_C(\Delta t)\>$ at the lowest investigated temperature in simulations,
clarifying that the average cage size changes in time significantly. 
At relatively small $\Delta t$, a smooth but clear grow of $\<S^2_C(\Delta t)\>$ takes place as $\Delta t$ vanishes,
which is the signature of a cage-opening process preceding a jump.
For $\Delta t\gtrsim100$, this trend becomes less marked, but survives up to long time,
since $\<S^2_C(\Delta t)\>$ monotonically increases as $\Delta t$ decreases, with a seemingly logarithmic behaviour.
This is a quite counter-intuitive result, as the cage dynamics is expected to be stationary at least well before a jump starts, 
and, therefore, should likely lead to a long-time plateau in $\<S^2_C(\Delta t)\>$.
To rationalize this result and reconcile it with the standard cage picture,
we consider that the cage duration is characterized by a broad distribution, $P(t_w)$, also known as waiting time distribution,
which decays as $t_w$ increases~\cite{SM14, JSTAT16, SM15}.
Since the value of $\<S^2_C(\Delta t)\>$ is only determined by those cages
of duration $t_w>\Delta t$, we speculate that the behavior of $\<S^2_C(\Delta t)\>$
could be explained assuming cages with different $t_w$ to have a different dynamics.
To confirm this hypothesis, we compute the average size, $\<S^2_C(\Delta t)\>_{t_w}$, over sub-ensembles of cages with a fixed lifetime $t_w$ 
(to improve statistics, $\<S^2_C(\Delta t)\>_{t_w}$ is operatively computed
as the average over all cages with waiting time in the range
$t\in[t_w-t_w/10, t_w+t_w/10]$).
Figure~\ref{fig:cage_t_tw}b compares $\<S^2_C(\Delta t)\>_{t_w}$, for a number of $t_w$, with the total average $\<S^2_C(\Delta t)\>$.  
At small $\Delta t$, curves corresponding to different $t_w$ overlap, 
indicating that the cage opening process is not affected by $t_w$.
Away from the jump, instead, results do depend on the cage duration.
Indeed, $\<S^2_C(\Delta t)\>_{t_w}$ attains a roughly $\Delta t$ independent value, $S^2_{t_w}$,
which is smaller for larger $t_w$. 
The upper inset of Fig.~\ref{fig:cage_t_tw}a shows the estimated values of $S^2_{t_w}$ as a function of $t_w$,
suggesting that the data are compatible with a power law plus a constant term,
$S^2_{t_w} = A/t_w^\beta + S^2_{\infty}$. 
 
We have performed the same analysis for the experimental system, 
finding fully consistent results as in Fig.~\ref{fig:cage_t_tw}c and d, 
although the sub-ensemble averages $\<S^2_C(\Delta t)\>_{t_w}$
are much more noise, due to poorer statistics.

Overall, these results demonstrate the existence of a coupling between cage size and duration.
In particular, one can assume that, for large $\Delta t$, 
the cage size, $\<S^2_C(\Delta t)\>_{t_w}$,  acquires a constant value determined by the overall cage duration, $t_w$.
These results lead to a simple model to rationalize the apparently anomalous behaviour of the average cage size, 
$\<S^2_C(\Delta t)\>$, at large $\Delta t$. Indeed, the average over the whole
ensemble of cages can be written as weighted sum over sub-ensemble averages:
\begin{equation}
\label{eq:S2_medio}
\<S^2_C(\Delta t)\>= \frac { \sum_{t_w=\Delta t}^{\infty}  \Omega(t_w) \<S^2_C(\Delta t)\>_{t_w} } {\sum_{t_w=\Delta t}^{\infty}  \Omega(t_w)}
\end{equation} 

where $\Omega(t_w)$ is the number of detected cages of lifetime $t_w$. 
Considering that this number is  proportional to the waiting time distribution, $\Omega(t_w) \propto P(t_w)$, and 
that for large $\Delta t$, $\<S^2_C(\Delta t)\>_{t_w}\simeq S^2_{t_w}$, Eq.\ref{eq:S2_medio} finally reads:

\begin{equation}
\label{eq:final}
\<S^2_C(\Delta t)\>=\frac{\int_{t_w=\Delta t}^{\infty} P(t_w) S^2_{t_w}  dt_w } {\int_{t_w=\Delta t}^{\infty} P(t_w) dt_w }
\end{equation}

which relates the cage size and the waiting time distribution for $\Delta t \gg0$.
In order to test this theoretical prediction, we have evaluated the r.h.s. of Eq.~\ref{eq:final} using our estimation 
for $S^2_{t_w}$ and $P(t_w)$ as a function of $t_w$.  
The prediction, reported as a dashed line in Fig.~\ref{fig:cage_t_tw}a and Fig.~\ref{fig:cage_t_tw}c,
describes very well the data for $\Delta t$ larger than the cage opening process.
Overall, these results suggest that the cage size only increases shortly before
a jump occurs, whereas the apparent long-time grow of $\<S^2_C(\Delta t)\>$ is the consequence
of averaging over an ensemble characterized by a coupling between cage size and duration, as well as, by a broad distribution of cage duration.

In this paper, we have investigated the single particle motion in \textgx{glass--forming liquids} 
to illuminate the relation between short-time dynamics of localized particles
and rearrangements leading to the structural relaxation on much longer timescales.
\textgx{The novel strategy we have introduced and the fundamental nature of the considered models
make this analysis directly applicable to a wide variety of biological, chemical and physical systems, in which particle crowding plays a major role.}
In particular, we focused on a nearly instantaneous dynamical property, the cage size, and explored its temporal evolution, 
as well as, the correlation with the following jumps.
Through the investigation of a jump propensity, we have clarified that particles rattling in large cages 
are more likely to jump after a short delay than particles rattling in small cages.   
Accordingly, the process of cage opening consists, on average, in a smooth enlargement of the cage, which lasts over a short time interval preceding the jump.
However, the correlation between jump propensity and cage size is only statistical, 
and progressively weakens as the time-lag between cage measurement and jump detection increases.
We provided evidences that, at large time-lags, the cage size is essentially constant and is related to the overall duration of the cage itself,
the smaller the cage the the longer the cage duration.
This coupling between cage size and duration suggests that cage size and local structural order are also intimately related.
For example, in the model numerically investigated in this paper, the time a particle spends in its cage before jumping is found to be correlated
with the local hexatic order~\cite{JSTAT16} and, therefore, similar correlations should in turn exist between the cage size and the hexatic order.
Accordingly, the higher the local order, the smaller the cage size and the longer the cage duration.
A possible explanation for this effect is that particles trapped in smalle cages are those packed in the core of highly ordered regions. 
Particles in the core of these regions can jump only when reached by a diffusing structural defect and, therefore, after those of the periphery,
resulting in larger cage duration. One such mechanisms has been reported for experiments on charged colloidal suspensions~\cite{Murray} and simulations of
glass-forming liquids~\cite{Mitus}. 

A further remark concerns the impressive similarity of the reported results, \textgx{that we have obtained by numerically investigating a molecular supercooled liquids 
and by experimentally studying a hard-sphere colloidal suspension. 
Such a similarity supports the universal nature the cage--jump motion~\cite{Chauduri, JSTAT16} and cannot be simply rationalized by the known dynamic scaling between hard and soft sphere systems, since 
this scaling is applicable to the family of soft potentials with inverse power--law dependence on the interparticle distance, $V(r)\propto 1/r^{\nu}$
(whose $\nu \rightarrow \infty$ limit corresponds to the hard sphere potential).~\cite{Medina, Medina_epl, Medina_pre, Dyre} 
This kind of dynamic equivalence does not hold for our supercooled liquid model, which is, instead, characterized by  a soft harmonic potential, not diverging for
$r\rightarrow 0$ and showing  properties, that cannot be mapped on hard-sphere-systems.~\cite{Max}.
}


While in this paper we investigate the correlation between the cage size and the first following jump, a possible extension of this work 
consists in considering subsequent jumps, that is, whether particles which have a large propensity to make a jump
are also likely to make many subsequent jumps. This is a way to investigate the life-time of dynamic heterogeneity from a single particle perspective~\cite{Kaufman}.
A closely related question is understanding to what extent it is possible to predict the jumps through measurement of the cage size at a previous time.
Similar studies have been performed in numerical works using the isoconfigurational ensemble~\cite{Widmer-Cooper, Candelier}, but the possibility
to make practical prediction in experimental systems is still an open issue.   

\section{Methods}
\label{sec:methods}
{\it Experimental.}
We have experimentally investigated a popular model system of hard-sphere colloidal \textgx{glass--forming} supension~\cite{Sood_NatComm, SM15, Sood_NatPhys, Sood_PRL, Vivek}.
Precisely, the sample consists in a 50:50 binary mixture of silica beads dispersed in water, at nearly monolayer condition. 
Bead diameters are $3.16\pm0.08$ and $2.31\pm0.03 ~\mathrm{\mu}m$ respectively,
resulting in a $\approx 1.4$ ratio known to prevent crystallization.
We image the system using a standard microscope equipped with a 40x objective
(Olympus UPLAPO 40XS). The images were recorded using a fast digital camera
(Prosilica GE680).
Particle tracking was performed using custom programs.
We have investigated different volume fractions $\phi$, in the range $0.64$--$0.79$,
where the systems can be properly equilibrated.
Increasing the volume fraction, the relaxation time, $\tau$, measured on the typical jump length and at thermal equilibrium,
increases in the interval   $10^2 s \gtrsim \tau \gtrsim 3~10^4 s$ and is compatible with a power-law 
functional form, $\tau(\phi) \propto (\phi_c-\phi)^{-c}$, with $\phi_c \simeq 0.81\pm0.01$
and $c = 2.6\pm0.02$. Further details on the systems and on the experimental set-up can be found in Ref.~\cite{SM15}. 

{\it Numerical simulations.}
We have performed NVT molecular dynamics 
simulations~\cite{LAMMPS} of a popular glass-forming model~\cite{Likos}.
The system consists in a two-dimensional 
50:50 binary mixture of $2N = 10^3$ disks,
with a diameter ratio $\sigma_{L}/\sigma_{S} =1.4$,
known to inhibit crystallization, at a fixed area fraction $\phi = 1$
in a box of side $L$. 
Particles interact via a soft potential, $V(r_{ij}) = \epsilon 
\left((\sigma_{ij}-r_{ij}) 
/\sigma_L\right)^\alpha \Theta(\sigma_{ij}-r_{ij})$, with $\alpha=2$ (Harmonic).
Here $r_{ij}$ is the inter-particle separation and $\sigma_{ij}$ the average 
diameter of the interacting particles.
This interaction and its variants (characterized by different values of $\alpha$)
are largely used to model dense colloidal systems,
such as foams~\cite{Durian}, microgels~\cite{Zaccarelli} and molecular glasses~\cite{Berthier_Witten, Manning}.
Units are reduced so that $\sigma_{L}=m=\epsilon=k_B=1$, 
where $m$ is the mass of both particle species and $k_B$ the Boltzmann's 
constant. The two species behave in a qualitatively analogous way,
and all data presented here refer to the smallest component.
In our simulations, the glass transition is approached by lowering the temperature 
in the range  $T \in [1.7~10^{-3}, 2.5~10^{-3}]$. At each investigated temperature, we monitor the dynamics after
fully equilibrating the systems. 
Lowering $T$, the relaxation time, $\tau$, increases in the interval   $2~10^2  \gtrsim \tau \gtrsim 10^4 s$ and is 
well described by a super-Arrhenius, $\tau(T) \propto \exp\left(A/T^2\right)$, with $A=const$~\cite{SM15_corr}.  

{\it Cage-jump detection algorithm.}
The trajectory of each particle is segmented in a series of cages
interrupted by jumps using an algorithm introduced in Ref.~\cite{SM14},
and largely tested both in simulations~\cite{SciRep, SM15_corr, JSTAT16} and experiments~\cite{SM15}.
To identify caged and jumping particles the algorithm compares the fluctuations of the 
particle position at a given time to the Debye-Waller factor.
To this end, we associate to each particle, at each time $t$,
the fluctuations of its position, $S^2(t)=\<r^2(t_0)\>-\<r(t_0)\>^2$, averaged
over the time interval $[t-\delta t/2, t+\delta t /2]$.
Following Ref.~\cite{Leporini} we defined the Debye-Waller factor
from the mean square displacement as $\<u^2\>=\<r^2(t_{DW})\>$, 
$t_{DW}$ being the time where $d\log (\<r^2(t)\>)/d\log (t))$ is minimal.
$\delta t$ is chosen of the order of 
$t_{DW}$\textgx{, therefore being much smaller than the relaxation time, $\tau$, but large enough for a particle
to experiment several collisions with its neighbours.
Accordingly, the algorithm is not sensitive to the single oscillation dynamics,
which is ballistic for molecular liquids and diffusive for colloidal suspensions.}  
Specifically, we use $\delta t =40$ for simulations and
$\delta t \simeq 60 s$ for experiments, respectively.
At time $t$, a particle is considered
in a cage if $S^2(t) < \<u^2\>$, and jumping otherwise.
When $S^2$ equals $\<u^2\>$, a particle is either starting or ending a jump/cage.
This algorithm allows for easily estimating 
the cage duration, $\tw$, which is the time-lag between two subsequent jumps of the same particle.
For a caged particle, a nearly instantaneous measure  of the size of  its cage at time $t$ 
 is, by construction, the fluctuation of the position at that time, $S^2_C(t)=S^2(t)$.

\begin{acknowledgement}

We acknowledge financial support 
from MIUR-FIRB RBFR081IUK, 
from the SPIN SEED 2014 project {\it Charge separation and charge transport in hybrid solar cells}
and from the CNR--NTU joint laboratory {\it Amorphous materials for energy harvesting applications}.
MPC acknowledge financial support from  the
Singapore Ministry of Education Academic Research Fund Tier 1 grants RG 104/15 and and RG 179/15.  

\end{acknowledgement}

\end{document}